\documentclass[aps,prb,superscriptaddress,twocolumn,showpacs]{revtex4-1}
\usepackage{amsmath,amssymb,mathrsfs,bbold}
\usepackage[dvips]{graphicx}

\newcommand{\bs}[1]{\boldsymbol{#1}}
\newcommand{\ii}{\text{i}}

\allowdisplaybreaks[4]

\begin{document}
%%%%%%%%%%%%%%%%%%%%%%%%%%%%%%%%%%%%%%%%%%%%%%%%
\title{Non-equilibrium transport through quantum dots with\\
  Dzyaloshinsky--Moriya--Kondo interaction}
%%%%%%%%%%%%%%%%%%%%%%%%%%%%%%%%%%%%%%%%%%%%%%%%
\author{Mikhail Pletyukhov}
\author{Dirk Schuricht}
\affiliation{Institute for Theory of Statistical Physics, 
RWTH Aachen, 52056 Aachen, Germany} 
\affiliation{JARA-Fundamentals of Future Information Technology}
 \date{\today}
\pagestyle{plain}

\begin{abstract}
  We study non-equilibrium transport through a single-orbital Anderson
  model in a magnetic field with spin-dependent hopping amplitudes. In the
  cotunneling regime it is described by an effective spin-1/2 dot
  with a Dzyaloshinsky--Moriya--Kondo (DMK) interaction between the spin
  on the dot and the electron spins in the leads. Using a real-time
  renormalization group technique we show that at low temperatures 
  (i) the DMK interaction is strongly renormalized, (ii) the renormalized
  magnetic field acquires a linear voltage dependence, and (iii) the
  differential conductance exhibits a voltage asymmetry which is
  strongly enhanced by logarithmic corrections. We propose
  transport measurements in which these signatures can be observed.
\end{abstract}
%05.10.Cc	Renormalization group methods
%73.21.-b 	Electron states and collective excitations in multilayers, quantum wells, mesoscopic, and  nanoscale systems
%73.21.La	Quantum dots
%73.63.-b		Electronic transport in nanoscale materials and structures
%73.63.Kv	Quantum dots

\maketitle

%%%%%%%%%%%%%%%%%%%%%%%%%%%%

\emph{Introduction.}---Over the past decade it has been established that 
electronic transport measurements through quantum dots and single molecules can be used 
to probe various coherent spin phenomena.~\cite{Hanson,Katsaros-10}
One of the simplest models for the theoretical description of quantum dots 
is the single-orbital Anderson model, which consists of an energy level
that can be occupied by up to two electrons. This energy level is coupled to
electronic reservoirs with hopping amplitudes $t^\alpha$ where $\alpha=\mathrm{L,R}$.
In the regime of suppressed charge fluctuations the system can be described by the Kondo 
model, in which the effective spin $\bs{S}$ on the dot is coupled to the electron 
spins $\bs{s}$ in the leads via a Heisenberg exchange interaction 
$\propto\bs{S}\cdot\bs{s}$ resulting in rich Kondo physics.~\cite{Kondo}
When the attached reservoirs are held at different chemical potentials, 
electrons are transported through the Kondo dot by cotunneling processes.
The complicated interplay between Kondo physics and the non-equilibrium processes
has led to the development of various renormalization group (RG) 
approaches.~\cite{Rosch,SchoellerReininghaus09} 

An interesting extension of the Anderson model is obtained by allowing spin-dependent
hopping amplitudes $t^\alpha_{\sigma\sigma'}$ where $\sigma,\sigma'=\uparrow,\downarrow$.
As it was shown by Paaske \emph{et al.}~\cite{Paaske-10} spin-orbit interaction in 
materials like SiGe may lead to a mixing of different orbital states of the quantum dot, thus 
resulting in an effective spin dependence of the hopping amplitudes. 
Observable manifestations of this effect are 
voltage asymmetries of the differential conductance~\cite{Paaske-10,Katsaros-10} and the 
suppression of Kondo ridges at finite magnetic fields.~\cite{Grap} 

However, spin-orbit interaction is not the only possibility to generate a spin dependence
of the hopping amplitudes. Pustilnik \emph{et al.}~\cite{Pust} have considered quantum dots with 
an even number of electrons in a finite magnetic field. Tuning the Zeeman energy to the
value of an orbital splitting creates a pair of degenerate levels, which 
originate from different orbital states and thus naturally possess different hopping 
amplitudes. They further derived the effective Kondo model for the case of 
symmetric hoppings to the leads, which possesses an anisotropic exchange 
interaction $\propto S^xs^x+S^ys^y+\Delta S^zs^z$.

\begin{figure}[b]
  \centering
  \includegraphics[width=60mm]{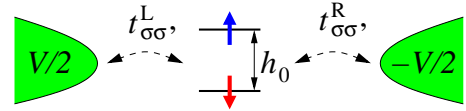}
  \caption{(Color online) Single-orbital Anderson model in a magnetic field 
        $\bs{h}_0$. The energies of the singly occupied levels are 
        $\epsilon_\mathrm{d}\pm h_0$. The dot levels are coupled to the leads 
        via spin-dependent hopping amplitudes 
        $t_{\sigma \sigma'}^\alpha = t^1_\alpha-t^2_\alpha (\bs{d} \cdot 
        \boldsymbol{\sigma})_{\sigma \sigma'}$. In the Kondo regime of strong 
        Coulomb repulsions $U$ with $\epsilon_\mathrm{d}=-U/2$ the system 
        is mapped onto the effective model Eq.~\eqref{eq:ham}.}
  \label{fig:model}
\end{figure}

In this Rapid Communication we study a single-orbital Anderson model coupled to electronic 
leads via arbitrary spin- and lead-dependent hopping amplitudes 
$t^{\alpha}_{\sigma\sigma'}$ (see Fig.~\ref{fig:model}). We demonstrate that with 
these two prerequisites the regime of single electron occupancy of the dot is 
generally described by an effective Kondo model with anisotropic exchange interaction
\emph{and a Dzyaloshinsky--Moriya--Kondo (DMK) interaction} $\propto
\bs{d} \cdot (\bs{S} \times \bs{s})$ between the effective spin on the 
dot and the electron spins in the leads. 

Dzyaloshinsky--Moriya (DM) interactions arise in most elementary form in
magnetic molecules.~\cite{Miyahara} For example, dimer molecules can
be modeled by two local spins coupled by a DM term, which can
give rise to a deviation of the orientation of the magnetization from the direction
of the external magnetic field. This situation is of particular relevance~\cite{SMM} for 
single-molecule magnets like Mn$_6$ and Mn$_{12}$. The DM 
interaction results in a canting of the spins that is also predicted to have 
characteristic signatures in the transport properties.~\cite{Herzog}

In sharp contrast to these previous works, the DMK interaction studied here 
does \emph{not} act within the quantum dot (or molecule) but rather between the spin 
(or pseudospin) on the dot and the electron spins in the leads. 
As the DMK interaction is antisymmetric under a left-right inversion of the system, it has 
drastic consequences for its non-linear transport properties, which we study using a
real-time renormalization group (RTRG) method.~\cite{Schoeller09} 
We find that (i) the DMK interaction is strongly renormalized during the RG flow, 
(ii) the renormalized magnetic field (RMF) $\bs{h}$ acquires a component along the 
DMK vector $\bs{d}$ that grows linearly with the bias voltage $V$, 
and (iii) the DMK interaction results in
an asymmetry of the differential conductance $G (V) \neq G (-V)$, which
is strongly enhanced by logarithmic corrections. 
We further discuss how the RMF manifests itself  in
the correlation functions and the real-time dynamics of observables, and 
propose the measurement of the differential conductance in two- and
three-terminal set-ups for its experimental detection.

\emph{Model.}---We consider a quantum dot coupled to two
electronic reservoirs held at chemical potentials $\mu_{\mathrm{L,R}}
= \pm V/2$ (Fig.~\ref{fig:model}). Electronic transport through the
dot is mediated by two levels with level splitting $h_0$, which are assumed to
be well separated from all other energy levels of the dot. 
The hopping amplitudes $t^\alpha_{\sigma \sigma'}$ depend on the 
spin states as well as the attached leads.
In the case of normal Fermi-liquid leads the
matrices $t^\alpha_{\sigma \sigma'}$ can be chosen hermitian and
thus parameterized as $t^\alpha_{\sigma \sigma'} = t^1_\alpha
\delta_{\sigma \sigma'} -t^2_\alpha (\bs{d}_\alpha \cdot
\boldsymbol{\sigma})_{\sigma\sigma'}$ in terms of four real numbers
$t^{1,2}_\alpha$ (one can always choose $t^1_{\alpha} >0$) and two unit vectors
$\bs{d}_{\alpha}$; $\boldsymbol{\sigma}$ denotes the vector
of Pauli matrices. We focus on the case 
$\bs{d}_{\alpha} \equiv \bs{d}=(d^i)$ where $i=x,y,z$. 

In the Kondo regime one can eliminate the empty and doubly occupied states on the dot via
Schrieffer-Wolff transformation.~\cite{Paaske-10} This yields an 
effective Kondo-type Hamiltonian
\begin{equation}
  \begin{split}
  \hspace{0mm}H=&\sum_{\alpha k\sigma}(\epsilon_k-\mu_\alpha)\,
  c_{\alpha k\sigma}^\dagger c_{\alpha k\sigma}+\sum_i h^i_0\,S^i \\
  &+\sum_{\alpha\beta ij}J_{\alpha\beta}^{ij}S^i\,s^j_{\alpha\beta}
  -\frac{1}{2}\sum_{\alpha\beta i}\bar{J}_{\alpha\beta}\,\, d^i \, S^i\,n_{\alpha\beta}, 
  \end{split}
  \label{eq:ham}
\end{equation}
where $\bar{J}_{\alpha\beta}= 8 (t_\alpha^1 t_\beta^2+t_\alpha^2
t_\beta^1)/U$ is the coupling between the spin-1/2 
on the dot and the charge density $n_{\alpha\beta}=\sum_{kk'\sigma} c_{\alpha
  k\sigma}^\dagger c_{\beta k'\sigma}$ in the leads, and $U$ denotes the on-site
Coulomb repulsion. The exchange coupling between $S^i$ and the spin densities
$s_{\alpha\beta}^i=\frac{1}{2}\sum_{kk'\sigma\sigma'} c_{\alpha
  k\sigma}^\dagger \sigma^i_{\sigma\sigma'}c_{\beta k'\sigma'}$ in the
leads is given by
\begin{equation}
 J_{\alpha\beta}^{ij}=
  (\delta_{ij} - d^i d^j)J^\perp_{\alpha\beta}+ d^i d^j J^d_{\alpha\beta} +
  \ii \sum_k \epsilon_{ijk} d^k J^\mathrm{DMK}_{\alpha\beta}.
  \label{eq:Jij}
\end{equation}
The couplings $J^{\perp/d}_{\alpha\beta}= 8 (t_\alpha^1
t_\beta^1\mp t_\alpha^2 t_\beta^2)/U$ constitute the anisotropic Kondo 
model,~\cite{SchoellerReininghaus09} while
$J^\mathrm{DMK}_{\alpha\beta}= 8 (t_\alpha^1 t_\beta^2-
t_\alpha^2 t_\beta^1)/U$ is  the Dzyaloshinsky--Moriya--Kondo (DMK) 
coupling between the spin on the dot and the electron spins in the leads.~\cite{2DEG}

An important parameter in the model \eqref{eq:ham} is the angle between
the DMK vector $\bs{d}$ and the applied magnetic
field $\bs{h}_0$. A parallel alignment $\bs{d}\parallel\bs{h}_0$ describes,
for example, the Kondo effect in quantum dots with an even number of electrons.~\cite{Pust}
On the other hand, the 
perpendicular case $\bs{d} \perp \bs{h}_0$ is
found~\cite{Paaske-10} in the effective cotunneling model for a
Kramers doublet in quantum dots with spin-orbit interactions, provided the coupling
between the orbital angular momentum and the magnetic field is neglected.

We stress that a non-vanishing DMK term requires a left-right asymmetry of
the hopping amplitudes, which is quite naturally present in nanoscale junctions,  
and contains the non-local reservoir spin
density $\boldsymbol{s}_\mathrm{LR}$. The corresponding coupling matrix
$J^\mathrm{DMK}$ is antisymmetric in reservoir space, i.e. $J^\mathrm{DMK}\propto
\ii\tau^y$. The other couplings $J^{\perp,d}$ and $\bar{J}$ are
symmetric and thus spanned by $\mathbb{1}$, $\tau^x$, and $\tau^z$, where
$\tau^i$ denote the Pauli matrices in reservoir space. Hence in total the model
\eqref{eq:ham} contains ten couplings. Although their
bare values are expressed just through the four parameters
$t^{1,2}_\mathrm{L,R}$, the relations between renormalized couplings
appear to be more complex, and all ten of them have to be taken
into account in the RG treatment.

\emph{Scaling analysis.}---The renormalization of the couplings
$J$ is governed to leading order by the poor-man's scaling (PMS)
equations. For the model \eqref{eq:ham} they have the form (in the
following all matrix operations are performed in the reservoir space)
\begin{equation}
  \begin{split}
  & \frac{\mathrm{d}}{\mathrm{d} l} J^\perp=
  \frac{1}{2}\big\{J^\perp,J^d\big\}
  +\frac{1}{2}\big[J^\mathrm{DMK},\bar{J}\,\big],\\
  & \frac{\mathrm{d}}{\mathrm{d} l} J^d=
  (J^\perp )^2 - (J^\mathrm{DMK})^2,\quad
  \frac{\mathrm{d}}{\mathrm{d} l} \bar{J}=
  \big[J^\perp,J^\mathrm{DMK}\big],\\
  & \frac{\mathrm{d}}{\mathrm{d} l} J^\mathrm{DMK}=
  \frac{1}{2}\big\{J^\mathrm{DMK},J^d\big\}
  +\frac{1}{2}\big[J^\perp,\bar{J}\,\big],
  \label{eq:PMS}
  \end{split}
\end{equation}
where $l = \ln (D/\Lambda)$ with $\Lambda$ being the flow parameter. Its
initial value is given by the band width $D$. The parametrization
\eqref{eq:Jij} is preserved under the RG flow governed by \eqref{eq:PMS}, i.e. no new
terms are generated. The PMS equations \eqref{eq:PMS} also leave
invariant $\mathrm{tr} \bar{J} = \mathrm{tr} \bar{J}_0
\equiv 2 c$ and $\mathrm{tr} \big[(J^\perp)^2-(J^d)^2-\bar{J}^2-
(J^\mathrm{DMK})^2\big]=0$. The latter relation implies that
when both $\bar{J}$ and $J^\mathrm{DMK}$ are initially zero one 
recovers the isotropic Kondo model with $J^\perp=J^d$.
We emphasize that despite $J^\perp$ and $J^\mathrm{DMK}$ formally satisfy the 
same RG equation, they have different symmetry properties in the reservoir space.

It is worthwhile to note that the magnetic field $\bs{h}_0$ does not explicitly enter the
PMS equations~\eqref{eq:PMS}. Its magnitude $h_0$ may appear as a
cut-off scale for the RG flow, but its direction is irrelevant for the PMS analysis.
Hence the PMS equations possess a higher symmetry than the full system \eqref{eq:ham}.
This allows us to arbitrarily choose the direction of $\bs{d}$, which is not affected by 
the RG flow. The Eqs.~\eqref{eq:PMS} with the initial (bare) values $J_0 = J
(\Lambda=D)$ stated after Eqs.~\eqref{eq:ham} and \eqref{eq:Jij} can
be solved exactly. Here we focus on the most significant features of the
RG flow and the role of the spin-charge and DMK terms.

The parameters $\mathrm{tr} J^d$ and $\mathrm{tr} J^\perp$, which
essentially determine the flow diagram of the model \eqref{eq:ham}, obey
\begin{equation}
\mathrm{tr} J^d = 2 c \frac{1+ (T_\mathrm{K} / \Lambda)^{4c}}{1-(T_\mathrm{K} /\Lambda)^{4c}}, \quad 
\mathrm{tr} J^\perp = \frac{4 c}{K} \frac{(T_\mathrm{K} / \Lambda)^{2c}}
{1-(T_\mathrm{K} /\Lambda)^{4c}},
\label{eq:PMSsol}
\end{equation}
where the invariant Kondo temperature is defined by $T_\mathrm{K} = 
D\,[(\mathrm{tr} J^d - 2 c)/(\mathrm{tr} J^d + 2 c)] ^{1/4c}$, and $c$
was defined above. In comparison with the
conventional anisotropic Kondo model,~\cite{SchoellerReininghaus09} the solution
\eqref{eq:PMSsol} additionally depends on the asymmetry parameter $K =
\sqrt{(\mathrm{tr} J_0^d)^2 - 4 c^2}/(\mathrm{tr} J_0^\perp)$, which
is in general different from unity unless both equalities $t^{1}_\mathrm{L}=
t^{1}_\mathrm{R}$ and $t^{2}_\mathrm{L}= t^{2}_\mathrm{R}$ hold. In the
latter case we have $J^\mathrm{DMK} =0$ and $\bar{J}=\mathrm{const.}$ 
during the RG flow, thus we recover the scaling equations of Ref.~\onlinecite{Pust}.

In contrast, in the generic case of broken left-right symmetry $J^\mathrm{DMK} $ and 
$\bar{J}$ are renormalized under the RG flow \eqref{eq:PMS} and acquire non-zero
values \emph{even if one of them is absent initially}. 
For example, we find $\mathrm{tr}[\ii J^\mathrm{DMK}\tau^y]
\propto\sqrt{K^2-1}\,\mathrm{tr} J^\perp$, i.e. the DMK coupling diverges for 
$\Lambda\to T_\mathrm{K}$. In the following we show that precisely $J^\mathrm{DMK} $ and 
$\bar{J}$ cause interesting observable effects. 

\emph{Renormalized magnetic field (RMF).}---One of these effects is the renormalization of the 
magnetic field. For its derivation we have to go beyond the PMS analysis.
We use the RTRG method,~\cite{Schoeller09,SchoellerReininghaus09} as 
it is particularly suited to analyze non-equilibrium effects. The derived flow equation for the RMF
can be integrated with the result
\begin{equation}
  \bs{h} =\left[1-\frac{1}{2}\mathrm{tr}
    (J^d_\mathrm{c}-J^d_0)\right] \bs{h}_0
  -\mathrm{tr}   \bigl[(\bar{J}_\mathrm{c}-\bar{J}_0)\tau^z\bigr]\,\frac{V}{2}
  \bs{d} .
 \label{eq:renh}
\end{equation}
The subscript $\mathrm{c}$ denotes the exchange couplings cut off at
the energy scale $\Lambda_\mathrm{c}\equiv\max\{h_0,V\} \gg T_\mathrm{K}$. The
first term in \eqref{eq:renh} is analogous to the first-order
result~\cite{MooreWen00} for the RMF in the anisotropic Kondo model.
The second term originates from the renormalization of the coupling 
between the dot spin and the electron density in the leads, which \emph{is generated
by the DMK interaction} [see Eq.~\eqref{eq:PMS}]. 
Hence the linear voltage dependence of \eqref{eq:renh} 
is a consequence of the left-right asymmetry and spin dependence of the hopping
amplitudes.  We stress that the voltage dependence already appears in
first order in the renormalized exchange couplings $J_\mathrm{c}$,
and that the effect on the magnitude $h$ of the RMF is \emph{maximal in 
the parallel case} $\bs{d}\parallel\bs{h}_0$.  The voltage dependence
implies that a RMF will be generated even in the absence of an external
field $\bs{h}_0$ and that $h(V)\neq h(-V)$. This is in sharp contrast to the anisotropic Kondo
model~\cite{SchoellerReininghaus09}, where the voltage dependence only
enters through logarithmic corrections in $O(J_\mathrm{c}^2)$
and the RMF remains symmetric under inversion of the bias.

The result  \eqref{eq:renh} has an immediate consequence for
the dynamics~\cite{TE} $\langle S^x(t)\pm\text{i}S^y(t)\rangle\propto e^{\pm i h t} $ of the
transverse magnetization. In the parallel case
the precession frequency $h$ depends on the relative sign of $h_0$ and 
$V$, which can be switched externally. The comparison of the
two values of $h$ directly yields the
voltage-dependent term in \eqref{eq:renh}. Furthermore, the dynamical spin susceptibility
possesses~\cite{SS09} characteristic features at $\Omega=h,h\pm V$,
which can also be used to probe the RMF.
The experimental observation of the precession
frequency as well as the spin susceptibility does, however, require
time-resolved measurements on the quantum dot. Although time-resolved 
experiments have been previously performed,~\cite{Schleser-04}
it would be rather challenging to extract quantitative results for the RMF. 
Thus we propose two alternative
ways to observe the effect of the RMF in stationary quantities.

\begin{figure}[t]
  \centering
  \includegraphics[width=76mm]{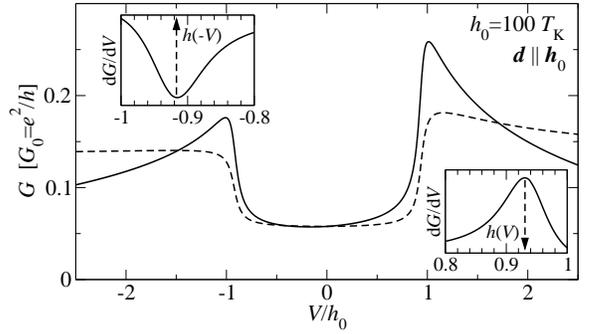}
  \caption{Differential conductance with (solid line) and without (dashed
    line) logarithmic corrections for $t_\mathrm{L}^1=0.08$, $t_\mathrm{L}^2=0.04$,
   $t_\mathrm{R}^1=0.05$, $t_\mathrm{R}^2=0.03$, $\bs{d}\parallel\bs{h}_0$, 
   and $h_0=100\,T_\mathrm{K}$. We observe an asymmetry under $V\to -V$, 
   which gets strongly enhanced by the logarithmic corrections. Insets: The 
   asymmetry of the RMF with respect to $V\to -V$ can be read off from the 
    resonances in $\mathrm{d}G/\mathrm{d}V=\mathrm{d}^2I/\mathrm{d}V^2$.}
  \label{fig:G}
\end{figure}
\emph{Differential conductance.}---As the DMK interaction breaks the inversion
symmetry of the system, it has drastic consequences on its non-equilibrium 
transport properties. In Fig.~\ref{fig:G} we plot the differential conductance $G (V)
=\mathrm{d}I/\mathrm{d}V$ through the dot for a parallel alignment of $\bs{d}$ and
$\bs{h}_0$. For $|V|<h$ the conductance is dominated by elastic cotunneling 
processes. At $V\approx \pm h$ the conductance possesses 
jumps due to the onset of inelastic cotunneling processes. 
The magnitude of these jumps is asymmetric with respect to $V\to -V$
already in order $J_\mathrm{c}^2$ (dashed line in Fig.~\ref{fig:G}), which 
qualitatively agrees with previous perturbative results.~\cite{Paaske-10} 
In leading order this asymmetry is proportional to 
$\mathrm{tr}[\ii J_\mathrm{c}^\mathrm{DMK}J_\mathrm{c}^\perp\tau^z] (\bs{d}\cdot\bs{h}_0)/h_0$. 
In addition, the RTRG analysis shows that the zero-temperature broadening of 
these jumps is given by the transverse spin relaxation rates $\Gamma_2(V=h)$ and
$\Gamma_2(V=- h)$, respectively, which differ from each other by
$\propto \mathrm{tr}\big[\ii J_\mathrm{c}^\mathrm{DMK}J_\mathrm{c}^\perp\tau^z\big]
(\bs{d}\cdot\bs{h}_0)$. This explicitly shows that the
asymmetry of $G (V)$ is a direct consequence of the DMK interaction and it can be
quantified in terms of the corresponding renormalized coupling
$J_\mathrm{c}^\mathrm{DMK} $. The asymmetry further depends on the angle
between $\bs{d}$ and $\bs{h}_0$, taking its maximal value in the
parallel case and vanishing in the perpendicular case.

A more thorough RTRG analysis in the parallel case including 
the leading logarithmic corrections in $O(J_\mathrm{c}^3)$ reveals that 
the magnitudes of the jumps at $V=\pm h$ acquire logarithmic terms $\propto
\ln(\Lambda_\mathrm{c}/ |V\mp h+\ii\Gamma_2(V\mp h)|)$ broadened by 
the corresponding transverse spin relaxation rates. Due to the difference 
of the respective prefactors the asymmetry of $G(V)$ acquires a strong enhancement
(see the solid line in Fig.~\ref{fig:G}), which even influences the elastic 
cotunneling regime $|V| < h$.

The logarithmic resonances at the onset of inelastic cotunneling at $V=\pm h$
correspond to the extrema of the derivative of the differential conductance 
$\mathrm{d}G/\mathrm{d}V=\mathrm{d}^2I/\mathrm{d}V^2$. Thus the RMF $h(\mp V)$
can be read off from the positions of its dip and peak respectively (see insets in 
Fig.~\ref{fig:G}). The second term in \eqref{eq:renh} is given by the 
sum of the two positions. We note that due to the restriction of our quantitative 
analysis to the weak-coupling regime the asymmetry $h(V)\neq h(-V)$  observed
in Fig.~\ref{fig:G} is relatively small. However, as it turns out to be a generic consequence 
of asymmetric and spin-dependent hopping amplitudes we expect it to be present in the
strong-coupling regime as well.

\begin{figure}[t]
  \centering
  \includegraphics[width=76mm,clip=true]{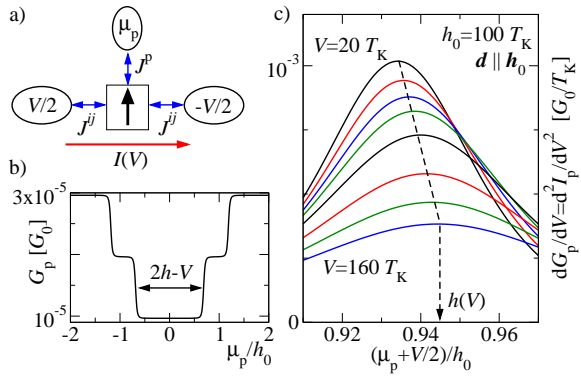}
  \caption{(Color online) a) Sketch of the three-terminal set-up. The
    stationary current $I(V)$ flows between the left and right leads
     while the probe lead is used to measure the RMF. b) Conductance 
    $G_\mathrm{p}$ in the probe lead for $J^\mathrm{p}=10^{-3}$ and $h_0=2V$. 
    All other parameters are as in Fig.~\ref{fig:G}. 
    c) The RMF \eqref{eq:renh} can be read off from the resonances in
    $\mathrm{d}/\mathrm{d}V\,[G_\mathrm{p}(\mu_\mathrm{p}+V/2)]$. The
    curves shown correspond (from top to bottom) to $V=20\,
    T_\mathrm{K},40\,T_\mathrm{K},\ldots,160\,T_\mathrm{K}$. The
    dashed line is a guide to the eye indicating the linear voltage
    dependence of $h(V)$.}
  \label{fig:plot1}
\end{figure}
\emph{Three-terminal set-up.}---Another way to experimentally observe the
linear voltage dependence of the RMF is provided by the use of the three-terminal set-up 
sketched in Fig.~\ref{fig:plot1}.a. A similar set-up has been used by Leturcq
\emph{et al.}~\cite{Leturcq-05} to measure the splitting of the Kondo
resonance in the spectral density. Specifically, the three-terminal set-up consists 
of the system \eqref{eq:ham} with an additional, weakly coupled probe lead modeled by
$\sum_{k\sigma}(\epsilon_k-\mu_\mathrm{p})\,c_{\mathrm{p}k\sigma}^\dagger
c_{\mathrm{p}k\sigma}$ with exchange interaction 
$J^\mathrm{p}\sum_{\alpha=\mathrm{L,R}}
\bs{S}\cdot(\bs{s}_{\alpha \mathrm{p}}+\bs{s}_{\mathrm{p}\alpha})
+ J^\mathrm{p}\,\bs{S}\cdot\bs{s}_{\mathrm{p}\mathrm{p}}$,
where $|J^\mathrm{p}|\ll|J^{ij}_{\alpha\beta}|,|\bar{J}_{\alpha\beta}|$. We 
assume symmetric and isotropic exchange couplings to the probe
lead and note in passing that the calculation can easily be generalized to more complicated
situations.  As the probe lead is only coupled weakly, its influence on the
PMS equations and the RMF is negligible, and the conductance
$G_\mathrm{p}=\mathrm{d}I_\mathrm{p}/\mathrm{d}\mu_\mathrm{p}$
can be calculated~\cite{SchoellerReininghaus09} using perturbation theory in $J^\mathrm{p}$.
The result is shown in Fig.~\ref{fig:plot1}.b: $G_\mathrm{p}$ possesses characteristic
steps at $\mu_\mathrm{p}=\pm (h-V/2),\pm(h+V/2)$ which are broadened by the
transverse spin relaxation rate. When plotting $\mathrm{d}G_\mathrm{p}/\mathrm{d}V$
as a function of $\mu_\mathrm{p}+V/2$ one can directly read off the RMF 
from the position of the first resonance. The results shown in 
Fig.~\ref{fig:plot1}.c clearly reveal the linear voltage dependence of the resonance 
position and thus the RMF.

\emph{Conclusion.}---We have studied a Kondo dot with additional DMK interaction, 
which is amplified during the RG flow. The DMK term results from the 
broken inversion symmetry of the system and manifests itself in 
the non-equilibrium transport properties. In particular, we showed that the RMF
acquires a component along the DMK vector depending linearly on the bias voltage, 
which can be experimentally detected in two- and three-terminal set-ups. 
In addition, the differential conductance becomes asymmetric with respect to the bias.
As these fingerprints of the DMK interaction rely on broken spin and spatial symmetry, 
they are also expected beyond the weak-coupling regime. 
A complete description of the strong-coupling regime remains, however, an open problem.

%%%%%%%%%%%%%%%%%%%%%%%%%%%%
We would like to thank S. Andergassen, H. Schoeller, and M. Wegewijs for very
valuable discussions.  This work was supported by the DFG through FG
723 (M.P.), FG 912, and the Emmy-Noether Program (D.S.).
%%%%%%%%%%%%%%%%%%%%%%%%%%%%


\begin{thebibliography}{10}

\bibitem{Hanson}
R. Hanson \emph{et al.},  Rev. Mod. Phys. \textbf{79}, 1217 (2007).

\bibitem{Katsaros-10}
G. Katsaros \emph{et al.}, Nat. Nanotechnol. {\bf 5}, 458 (2010).
%G. Katsaros, P. Spathis, M. Stoffel, F. Fournel, M. Mongillo, V. Bouchiat,
%F. Lefloch, A. Rastelli, O. G. Schmidt, and S. De Franceschi, 

\bibitem{Kondo}
W.~G. van~der Wiel \emph{et al.}, Science {\bf 289},  2105  (2000);
N. Roch \emph{et al.}, Nature {\bf 453}, 633 (2008).

\bibitem{Rosch}
A. Rosch \emph{et al.}, Phys. Rev. Lett. {\bf 90}, 076804  (2003);
S. Kehrein, \emph{ibid.} {\bf 95},  056602  (2005);
F.~B. Anders, \emph{ibid.} {\bf 101},  066804  (2008).

\bibitem{SchoellerReininghaus09}
H. Schoeller and F. Reininghaus, Phys. Rev.~B {\bf 80},  045117  (2009);
  \emph{ibid.} \textbf{80}, 209901(E) (2009).
  
\bibitem{Paaske-10}
J. Paaske, A. Andersen, and K. Flensberg, Phys. Rev. B {\bf 82},  081309(R) (2010).

\bibitem{Grap} 
S. Grap \emph{et al.}, Phys. Rev. B {\bf 83}, 115115 (2011).
%S. Grap, S. Andergassen, J. Paaske, and V. Meden,

\bibitem{Pust}
M. Pustilnik, Y. Avishai, and K. Kikoin, Phys. Rev. Lett. {\bf 84}, 1756 (2000).

\bibitem{Miyahara}
S. Miyahara \emph{et al.}, Phys. Rev. B {\bf 75} 184402 (2007).
%S. Miyahara, J.-B. Fouet, S. R. Manmana, R. M. Noack, H. Mayaffre, 
%I. Sheikin, C. Berthier, and F. Mila, Phys. Rev. B {\bf 75} 184402 (2007).

\bibitem{SMM}
S. Bahr  \emph{et al.}, Phys. Rev. B {\bf 78}, 132401 (2008); 
W. Wernsdorfer, T. C. Stamatatos, and G. Christou, Phys. Rev. Lett. \textbf{101}, 
237204 (2008).

\bibitem{Herzog}
S. Herzog and M. Wegewijs, Nanotechnology {\bf 21}, 274010 (2010).

\bibitem{Schoeller09}
H. Schoeller, Eur. Phys.~J. Special Topics {\bf 168},  179  (2009).

\bibitem{2DEG}
A similar exchange interaction is generated for impurities
in a two-dimensional electron gas with Rashba spin-orbit interactions away from the 
particle-hole symmetric point [M. Zarea, S. E. Ulloa, and N. Sandler, arXiv:1105.3522v1].

\bibitem{MooreWen00}
J.~E. Moore and X.-G. Wen, Phys. Rev. Lett. {\bf 85},  1722  (2000);
M. Garst  \emph{et al.}, Phys. Rev. B  {\bf 72},  205125  (2005).
%M. Garst, P. W\"olfle, L. Borda, J. von Delft, and L. Glazman, 

\bibitem{TE}
Corrections to the Markov approximation can be calculated following
M. Pletyukhov, D. Schuricht, and H. Schoeller, Phys. Rev. Lett. \textbf{104}, 
106801 (2010).

\bibitem{SS09}
D. Schuricht and H. Schoeller, Phys. Rev.~B {\bf 80},  075120  (2009);
P. Fritsch and S. Kehrein, \emph{ibid.} {\bf 81},  035113  (2010).

\bibitem{Schleser-04}
R. Schleser \emph{et al.}, Appl. Phys. Lett. {\bf 85},  2005  (2004);
P.-F. Braun \emph{et al.}, Phys. Rev. Lett. {\bf 94},  116601  (2005);
K.~C. Nowack \emph{et al.}, Science {\bf 318},  1430  (2007).

\bibitem{Leturcq-05}
R. Leturcq \emph{et al.}, Phys. Rev. Lett. {\bf 95},  126603  (2005).
%R. Leturcq, L. Schmid, K. Ensslin, Y. Meir, D.~C. Driscoll, and A.~C. Gossard,

\end{thebibliography}
\end{document}